\pdfoutput=1
\documentclass[texmf,cernpreprint,UKenglish,texlive=2011,txfonts]{atlasdoc}

\usepackage{atlaspackage}

\usepackage{atlasbiblatex}

\usepackage{atlasphysics}

\usepackage{graphicx}
\usepackage{placeins}

\AtlasTitle{A measurement of material in the ATLAS tracker using secondary hadronic interactions in 7 TeV \emph{pp} collisions}

\author{The ATLAS Collaboration}

\date{}

\PreprintIdNumber{CERN-EP-2016-137}

\AtlasJournalRef{JINST 11, (2016), P11020}
\AtlasDOI{10.1088/1748-0221/11/11/P11020}

\AtlasAbstract{%
 Knowledge of the material in the ATLAS inner tracking detector is crucial in understanding the reconstruction of charged-particle tracks, the performance of algorithms that identify jets containing \emph{b}-hadrons and is also essential to reduce background in searches for exotic particles that can decay within the inner detector volume. Interactions of primary hadrons produced in \emph{pp} collisions with the material in the inner detector are used to map the location and amount of this material. The hadronic interactions of primary particles may result in secondary vertices, which in this analysis are reconstructed by an inclusive vertex-finding algorithm. Data were collected using minimum-bias triggers by the ATLAS detector operating at the LHC during 2010 at centre-of-mass energy $\sqrt{s}$~=~7~\TeV, and correspond to an integrated luminosity of $19$~nb$^{-1}$. Kinematic properties of these secondary vertices are used to study the validity of the modelling of hadronic interactions in simulation. Secondary-vertex yields are compared between data and simulation over a volume of about 0.7~m$^3$ around the interaction point, and agreement is found within overall uncertainties. 
}

\begin{document}

\maketitle

\tableofcontents

\section{Introduction}

An accurate description of the amount and location of the material in the ATLAS inner detector (ID) is necessary for understanding the reconstruction efficiency of tracks, as well as other objects such as electrons, photons, and jets containing \emph{b}-hadrons. Moreover, uncertainties in the modelling of hadronic interactions in simulation affect the performance of algorithms that reconstruct jets and missing transverse momentum. The amount of material in the as-built ID~\cite{detector} is generally known to an accuracy of about 4--5\%, which was achieved by weighing the various components before installation.

Traditionally, photon conversions, sensitive to the radiation length ($X_0$), have been used to map the material in particle detectors. A complementary approach, based on hadronic interactions of primary particles with material in the ID, has been developed~\cite{VJain}, which is sensitive to the hadronic interaction length ($\lambda$). Secondary hadronic interactions with material in the ID usually involve low- to medium-energy primary hadrons with average momentum around 4 GeV, and with about 96\% having momentum less than 10~\GeV. Consequently, the outgoing particles have relatively low energy and large opening angles between them. The trajectories of these particles can be reconstructed by the tracking system, and in cases where two or more such tracks are reconstructed, the precise location of the interaction can be found.

In ATLAS, an inclusive vertex-finding and fitting algorithm is used to find vertices associated with hadronic interactions. Typical resolutions of vertex positions,\footnote{ATLAS uses a right-handed coordinate system with its origin at the nominal interaction point (IP) in the centre of the detector and the $z$-axis along the beam pipe. The $x$-axis points from the IP to the centre of the LHC ring, and the $y$-axis points upward. Cylindrical coordinates $(r,\phi)$ are used in the transverse plane, $\phi$ being the azimuthal angle around the beam pipe. The pseudorapidity is defined in terms of the polar angle $\theta$ as $\eta=-\ln\tan(\theta/2)$.} in \emph{r} (\emph{z}), are around 0.2~mm (0.3~mm) for this technique, as compared to around 2~mm (1~mm) for photon conversions. In the latter, the opening angle between the outgoing electron--positron pair is close to zero, thereby degrading the radial resolution. The good position resolution of hadronic-interaction vertices is exploited to study the location and amount of material in the ID.

Knowledge of the exact location of this material is needed in searches for exotic particles that can decay within the ID volume, as secondary interaction vertices are a background to these searches. The results from a previous study were used in analyses searching for decays of exotic particles that are postulated to exist in models ranging from R-parity-violating supersymmetry, split supersymmetry, Generalized Gauge Mediation~\cite{Aad:2015rba} to Hidden Valley scenarios~\cite{lubatti}. In these analyses, part of the search region was masked out using material maps based on the results of Ref.~\cite{VJain}. Additionally, since this technique directly probes the hadronic interaction length of any material, it can also be used to study the modelling of low- to medium-energy hadronic interactions in simulation.

The analysis presented here uses a modified tracking configuration and improved vertex selection criteria, which substantially increase the number and quality of reconstructed vertices over a larger volume as compared to the previous study. The previous study was restricted to $r<320$~mm and $|z|<300$~mm, whereas this one expands this region to $r < 400$~mm and $|z| < 700$~mm, which represents an increase in the fiducial volume from 0.19~m$^3$ to 0.70~m$^3$. The dataset used now is the same one as in Ref.~\cite{VJain}, although the track reconstruction algorithms use improved ID alignment parameters. The data are compared to simulation, which has an improved description of the ID geometry and other corrections to resolve some of the discrepancies found in the previous study.

This paper is structured as follows. A brief introduction to the ATLAS ID is given in Section~\ref{sec:ATLASID}. The event samples, track selection and vertex reconstruction and selection are described in Section~\ref{sec:Samples}, along with the yield of reconstructed secondary vertices (SV). Section~\ref{sec:QualComp} contains a qualitative comparison of data with simulation, followed by an estimate of the systematic uncertainties in Section~\ref{sec:syst}. The results are presented in Section~\ref{sec:Results}, and conclusions are given in Section~\ref{sec:Conclusions}.

\section{ATLAS inner detector}
\label{sec:ATLASID}
\begin{figure}[htbp]
\begin{center}
\includegraphics[width= 1.0\textwidth]{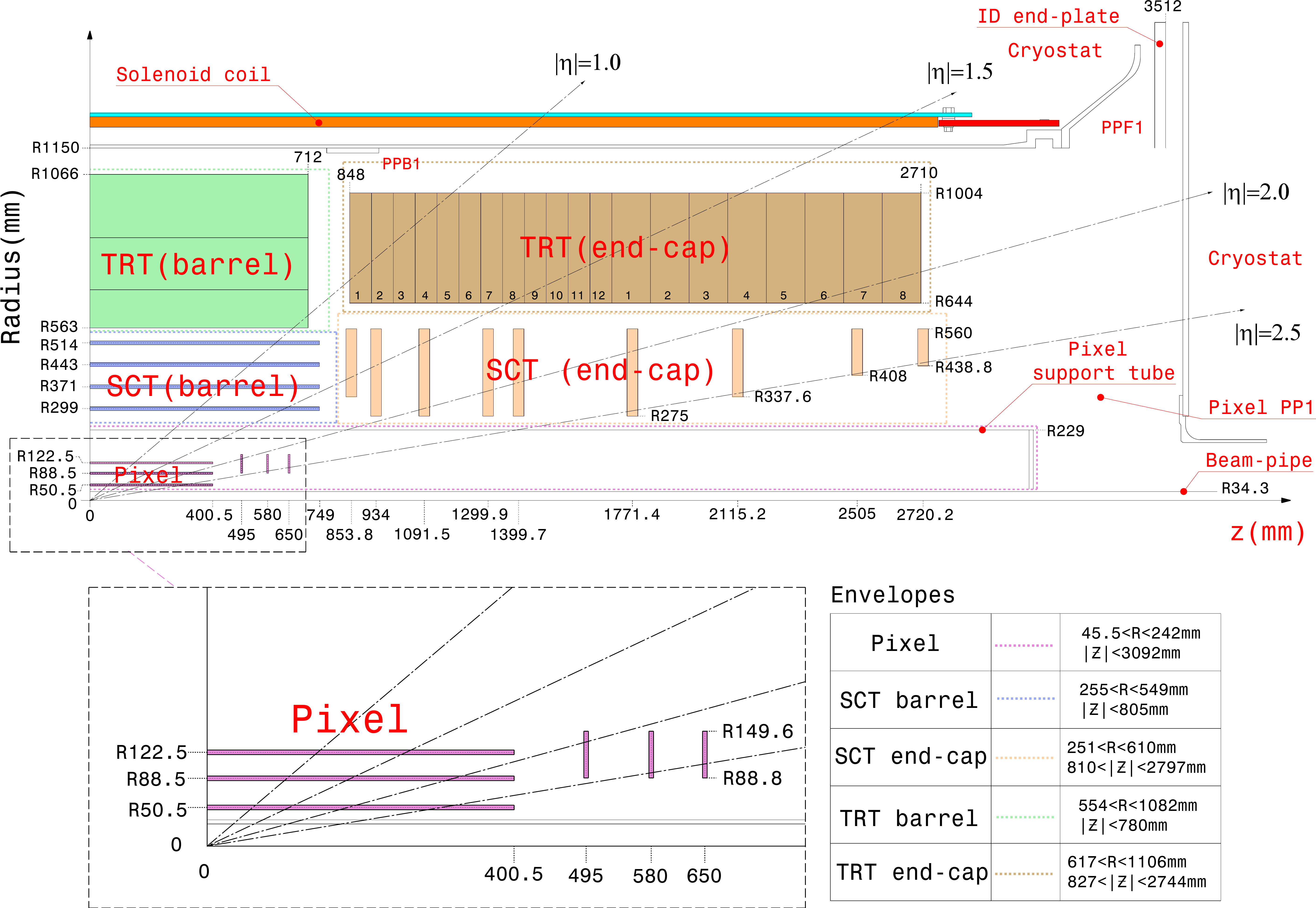}

\caption{View of a quadrant of the ID~\cite{Expect} showing each of the major detector elements with their active dimensions and envelopes. The lower part shows a zoomed-in view of the pixel detector region.
\label{fig:indet}}
\end{center}
\end{figure}

The ATLAS detector consists of an inner tracking detector, electromagnetic and hadron calorimeters, a muon spectrometer, and three magnet systems~\cite{detector}. The inner tracking detector is crucial to this analysis; its active volume extends from 45 mm to 1150 mm in \emph{r} and $\pm2710$~mm in $|z|$. A quarter section of the ID is shown in Figure~\ref{fig:indet}.

The ID is composed of three sub-systems: a silicon pixel detector, a silicon microstrip detector (SCT) and a transition radiation tracker (TRT), all of which are immersed in a 2~T axial magnetic field. The tracking system consists of a cylindrical barrel region ($|\eta| \lesssim 1.5$) arranged around the beam pipe, and two end-caps. Disks in the end-cap region are placed perpendicular to the beam axis, covering $1.5 \lesssim|\eta| < 2.5$. The pixel detector is located at $\emph{r}<150$ mm, and provides precision measurements from $80.4$ million sensors. It consists of three barrel layers with $|z|<400$ mm and six disks in the end-cap region, which extend out to $|z| \sim 660$~mm. The SCT detector has four barrel layers extending from $\emph{r} \sim 250$~mm to 550~mm with $|z|<750$ mm, and 18 disks in the end-cap region with $|z|<2710$ mm; it consists of $\sim6.3$ million readout strips. The TRT consists of 298,000 straw tubes with diameter 4 mm, and provides coverage out to $|\eta|\sim2.0$. In the barrel, it stretches from $\emph{r}\sim560$~mm to 1100~mm and $|z|<780$~mm. The end-cap region extends the TRT to $z = \pm2710$~mm. The regions between the material layers are filled with different gases, such as CO$_{2}$ or N$_2$. In addition, there are various support structures, which, for the most part, are carbon fibre reinforced plastic honeycomb shells, and services such as cooling pipes.

\section{Event samples, track and secondary-vertex reconstruction}
\label{sec:Samples}
Data were collected in 2010 \emph{pp} collisions at a centre-of-mass energy of 7$\TeV$ and correspond to an integrated luminosity of 19~nb$^{-1}$. Triggers used to collect data relied on the coincidence between the beam pickup timing devices ($z = \pm 175$~m) and minimum-bias trigger scintillators ($z = \pm 3.56$~m)~\cite{PrimTrk}. The instantaneous luminosity in the early part of the run was approximately $10^{27}$--$10^{29}$~cm$^{-2}$ s$^{-1}$, which implies, on average, a very low number of extra interactions per beam crossing (pile-up). This data sample is the same as in Ref.~\cite{VJain}, although the track reconstruction algorithms use improved ID alignment parameters. The data are compared to simulation, which includes an improved description of the ID geometry, e.g., the position of the beam pipe, the implementation of a slight shift of the pixel barrel layers, and the amount of material in the end-region of the pixel barrel.

Monte Carlo (MC) simulation events were generated using PYTHIA8~\cite{PYTHIA} with the MSTW2008LO set of parton distribution functions~\cite{PDF} and A2MSTW2008LO set of tuned parameters~\cite{tune}. Particle propagation through the detector was simulated with  GEANT4~\cite{GEANT4}.  Hadronic interactions in GEANT4 were simulated with the FTFP\_BERT~\cite{Models} model, which is an improvement over the previous model. The simulated events were processed with the same reconstruction software as data. The ATLAS simulation infrastructure is described in Ref.~\cite{atlasSimu}. 

The triggers that were used to collect the data sample recorded single-, double- and non-diffractive events, with a large fraction belonging to the last category. In order to reduce systematic uncertainties from single- and double-diffractive events, their contributions were reduced by requiring a large track multiplicity at the primary vertex (PV). Hence, an event was required to have exactly one reconstructed PV with at least 11 associated tracks~\cite{PVref}. After this requirement there were 18 (6) million events in data (simulation). The simulated events were weighted such that the mean and width of the \emph{z}-coordinate distribution of the PV position matched the data.

\subsection{Track reconstruction}
\label{sec:TrackReco}

Since tracks originating from secondary hadronic interactions generally have large impact parameters with respect to the PV, the reconstruction efficiency of such tracks needs to be as high as possible. In the standard track reconstruction algorithm, stringent upper limits are placed on the impact parameters of the tracks with respect to the beam axis. Such requirements have little effect on the efficiency of reconstructing primary tracks and speed up event reconstruction time. However, these requirements severely limit the reconstruction of secondary tracks, especially those originating far from the PV.

\begin{figure}[htbp]
\begin{center}
\begin{tabular}{cc}
\includegraphics[width=0.47\textwidth]{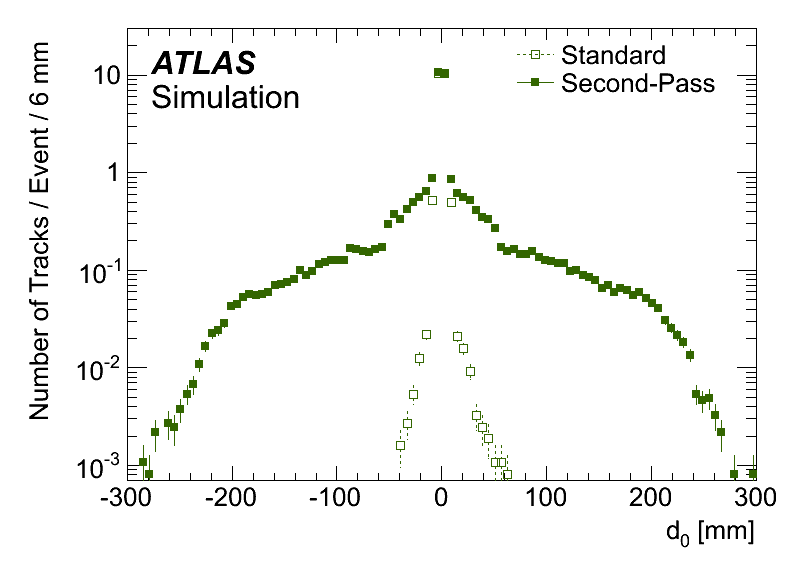} & \includegraphics[width=0.47\textwidth]{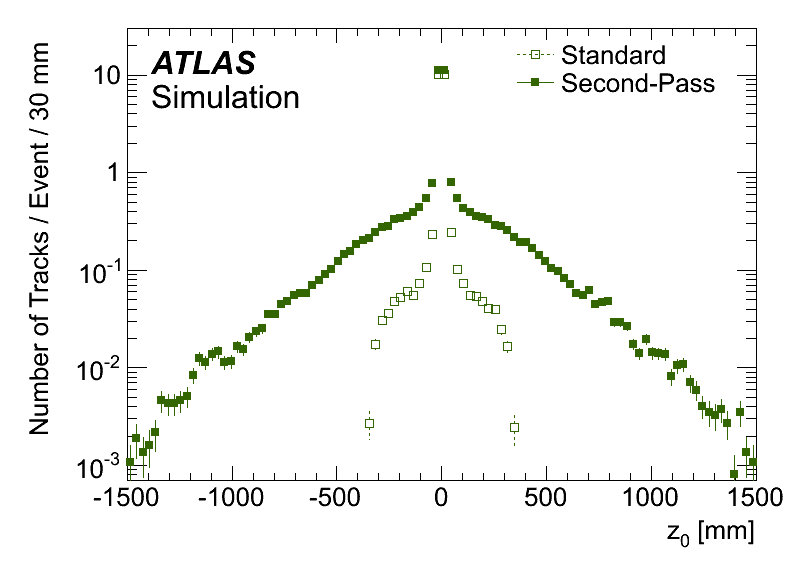} \\
(a) & (b) \\
\end{tabular}
\caption{Distributions of (a) transverse ($d_0$) and (b) longitudinal ($z_0$) impact parameters of tracks in simulated events. Open squares show tracks reconstructed with the standard algorithm, closed squares show tracks reconstructed with either the standard or second-pass reconstruction algorithm.
\label{fig:retrack}}
\end{center}
\end{figure}

To address the low efficiency of reconstructing secondary tracks, a second pass of track reconstruction was executed that used hits in the ID left over after the standard track reconstruction step had finished. This second pass also had looser selection criteria applied to track parameters.  The increase in the number of reconstructed tracks with large impact parameters can be seen in Figure~\ref{fig:retrack}, where $d_0$ ($z_0$) is the transverse (longitudinal) impact parameter with respect to the PV. In both reconstruction steps, the transverse momentum, \pt, of tracks was required to be $\ge 0.4$~\GeV, and their pseudorapidity was required to be within the range $|\eta| \le 2.5$.

\subsection{Track selection}
\label{sec:TrackSelection}

Secondary tracks typically have large values of transverse impact parameter. Therefore, in order to select mostly secondary tracks and reduce the contamination of primary tracks, the tracks were required to have $|d_0|>5$~mm. This requirement removed $\sim99\%$ of primary tracks as well as many tracks originating from $\Kshort$ and $\Lambda$ decays, and photon conversions.

No requirements were placed on the number of pixel hits on a track as this would have prevented reconstruction of vertices outside the pixel detector. However, tracks were required to have at least one SCT hit, to ensure that there was at least one hit per track in a silicon detector. In order to improve the quality of reconstructed secondary vertices, well-measured tracks were required, i.e., track fit $\chi^{2}/$dof~$< 5$ (where ``dof'' means degree of freedom).

Track reconstruction efficiency and track parameters are well reproduced in simulation. However, there are small differences between data and simulation ~\cite{PrimTrk, PYTHIA} in the number of reconstructed primary tracks. The primary track multiplicity directly affects the number of secondary interactions; if there are more primary particles, then there are also more hadronic interactions. Since data has more primary tracks, yields in simulation are corrected to account for this difference. This correction factor was determined by extrapolating primary tracks, separately in data and simulation, to various material layers, counting how many of them intersected the layer within the fiducial volume, and taking their ratio. Particles that travel through a layer at angles different than normal encounter more material and have a higher interaction probability than those at normal incidence; this is taken into account. The correction factor ranged from 1.05 to 1.07.

\subsection{Secondary-vertex reconstruction}
\label{sec:VtxRecoSel}

A \emph{pp} collision event may contain decays of short-lived particles such as bottom and charm hadrons, decays of long-lived particles such as $\Kshort$ and $\Lambda$, photon conversions, and hadronic interaction vertices with \emph{a~priori} unknown multiplicity. To cleanly detect these hadronic interactions, all other secondary vertices must be reconstructed and eliminated.

A vertex finder designed to simultaneously find all secondary vertices in the event was used in this analysis. The algorithm begins by finding all possible intersections of pairs of selected tracks, and assumes that the two secondary tracks originate from a single point. A fit is performed, during which track parameters are varied, and modified, if necessary, and the vertex position is determined. A $\chi^2$ describing the fit quality of the vertex is determined using the differences between the measured and the recalculated track parameters~\cite{vtxchi2}. The reconstructed two-track vertices define the full set of vertices in the event, because any \emph{N}-track vertex is simply a union of corresponding two-track sub-vertices.

Requiring these two-track vertices to have an acceptable $\chi^2$/dof ($<\!4.5$) removes $\sim\! 85\%$ of random pairings. Studies on simulated events indicate that more than 83\% of hadronic interaction vertices are retained. In the barrel region of the ID, the number of fake vertices from random combinations is further reduced by requiring that tracks do not have hits in silicon layers at a radius smaller than the radius of the reconstructed vertex, and have hits in some of the layers that are at larger radii than the vertex; in the end-cap region, only the minimum number of hits in the SCT is required. Vertices that fail to meet these criteria are removed from the list of selected two-track vertices. This procedure removes, depending on radius, anywhere from one-half to two-thirds of the initial set of two-track vertices, whereas the reduction in efficiency for reconstructing hadronic interaction vertices is about 2--10\%.

To finalize the vertex finding, two further steps were undertaken: first, the total number of vertices in the event was minimized by merging the two-track candidates that were nearby, a decision that was based on the separation between vertices and on their covariance matrices; second, since a track could have been used in several two-track vertices, such cases were identified and resolved so that all track--vertex associations were unique. The algorithm worked iteratively to clean the vertex set, based on an incompatibility-graph approach~\cite{IEEE}. In each iteration it either identified two close vertices and merged them, or found the worst track--vertex association for multiply assigned tracks and broke it. Iterations continued until no further improvement was possible. The CPU performance of the algorithm was acceptable for events with track multiplicity up to $\sim$~200, which was significantly larger than the average multiplicity in events used in this analysis (about 50 tracks/event).

\subsection{Secondary-vertex selection and resolution}
\label{sec:VtxQC}

The looser requirements on track reconstruction (Section~\ref{sec:TrackReco}) significantly improve vertex reconstruction efficiency for all values of \emph{r} and \emph{z}. However, studies on simulated events show that this increase is accompanied by a decrease in the purity of secondary-vertex reconstruction, which is defined as the fraction of reconstructed vertices that match true secondary interactions. The matching is based on the distance between the true and reconstructed position of a secondary interaction. Since the most important purpose of the analysis is to provide precise comparison of data and simulation, purity needs to be as high as possible. To achieve this, stringent requirements were placed on tracks associated with reconstructed secondary vertices: relative error on the transverse momentum of tracks $\sigma(p_{\mathrm T})/p_{\mathrm T}<0.05$, transverse and longitudinal impact parameters relative to the secondary vertex $|d_{0}^{\mathrm{SV}}|<1$~mm and $|z_{0}^{\mathrm{SV}}|<2$~mm respectively. 

These requirements were applied after track selection and vertex reconstruction since they were on quantities measured relative to the reconstructed secondary vertex. If a track in a secondary vertex failed these requirements, the entire vertex was discarded. This procedure removed a small fraction of real secondary vertices with more than two tracks, but since most of the reconstructed vertices had only two tracks, the benefit from the significant increase in purity outweighed the small loss in efficiency. In both data and simulation, about $90\%$ of reconstructed secondary vertices are two-track vertices and almost all of the other~$10\%$ of vertices have three tracks. Purities are consistently higher for all \emph{r} and \emph{z}, especially at larger radii, as compared to the previous study. For instance, the second SCT layer was not previously studied since the purity was only 15\%. Table~\ref{tab:TabPurity} lists the values at various material layers. This metric enters into the final result as an additional systematic uncertainty due to small differences in the rate of fake vertices in data and simulation. The total efficiency to find secondary vertices, containing at least two tracks with \pt~and $\eta$ satisfying the selection criteria, is about 9\% at the beam pipe, and decreases with radius; it includes efficiencies for track and vertex reconstruction steps, and the selection requirements described above.

\begin{table}
\begin{center}
\caption{\label{tab:TabPurity} Purity of reconstructed vertices in various regions of the ID. Errors are statistical only.}
\vskip 0.1cm
\begin{tabular}{|l|c|}
\hline
\multicolumn{2}{|c|}{Pixel barrel region ($|\emph{z}| < $400 mm)} \\
\hline \hline
Layer & Purity \% \\
\hline
Beam pipe (\emph{r}:28--36 mm) & $82.5\pm0.1$ \\
1st pixel layer (\emph{r}:45--75 mm) & $75.3\pm0.1$ \\
2nd pixel layer (\emph{r}:83--110 mm) & $80.0\pm0.1$ \\
3rd pixel layer (\emph{r}:118--145 mm) & $68.0\pm0.1$ \\
\hline \hline
\multicolumn{2}{|c|}{SCT barrel region ($|\emph{z}| < $700 mm)} \\
\hline \hline
Layer & Purity \% \\
\hline
Pixel support frame (\emph{r}:180--220 mm) &  $71.9\pm0.2$ \\
Pixel support tube (\emph{r}:226--240 mm) & $ 87.0 \pm0.4$ \\
1st SCT layer (\emph{r}:276--320 mm) & $78.1\pm0.2$ \\
2nd SCT layer (\emph{r}:347--390 mm) & $61.4\pm0.5$ \\
\hline \hline
\multicolumn{2}{|c|}{Pixel forward region (\emph{r}: 75--180 mm, $|\emph{z}| >$ 400 mm)} \\
\hline \hline
Layer & Purity \% \\
\hline
1st pixel disk  ($|\emph{z}|$:490--500 mm) & $57.7\pm0.4$ \\
2nd pixel disk ($|\emph{z}|$:575--585 mm) & $61.9\pm0.5$ \\
3rd pixel disk ($|\emph{z}|$:645--655 mm) & $56.7\pm0.7$ \\
\hline
\end{tabular}
\end{center}
\end{table}

The spatial resolution of the position of the secondary vertex depends on the quality of the track reconstruction. Vertices at smaller radius contain tracks with more silicon hits and are therefore reconstructed with less uncertainty. Averaging over the full kinematic range of tracks, studies on simulated events indicate that radial and \emph{z} resolutions at the beam pipe and through to the first SCT layer are about $0.22$~mm and $0.25$~mm respectively (see Ref.~\cite{VJain} for details). At the second SCT layer ($r > 320$~mm), they worsen to about $0.34$~mm and $0.70$~mm respectively. These resolutions make it possible to resolve structural details at the millimeter scale.

\subsection{Vertex yields}
\label{sec:recoData}

\begin{figure}[!htb]
\centering
\includegraphics[width=0.7\textwidth]{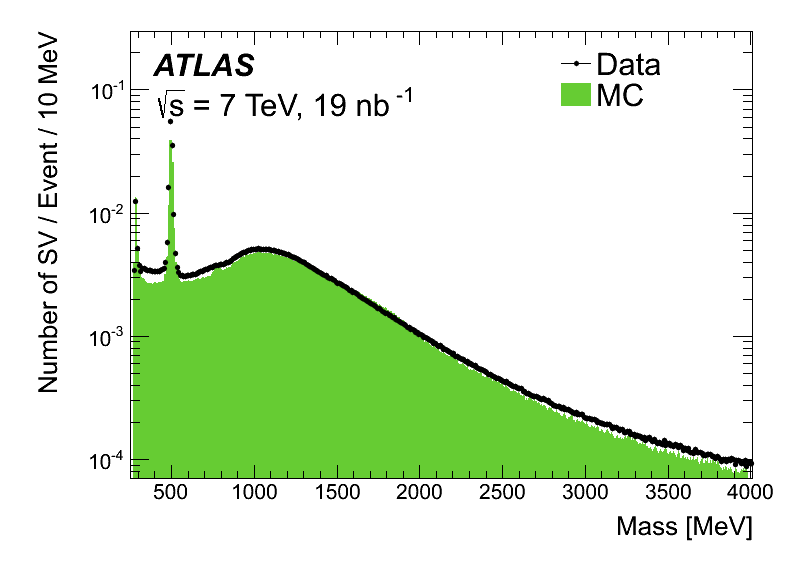}
\caption{Invariant-mass spectrum of the reconstructed secondary vertices, assuming the pion mass for all tracks, in data (points) and MC simulation (solid histogram).}
\label{fig:MassSpectra}
\end{figure}

The invariant-mass spectrum of reconstructed secondary vertices is shown in Figure~\ref{fig:MassSpectra}. During track reconstruction, the pion mass is assigned by default to tracks. The broad `shoulder' around $1100$~\MeV~is a kinematic effect and reflects the minimum required track \pt. The peak around $300~\MeV$~corresponds to photon conversions, where the non-zero mass is due to the assignment of the $\pi^{\pm}$ mass to the $e^{\pm}$ tracks, and the peak around $500~\MeV$~corresponds to $\Kshort$-meson decays. $\Lambda$-baryon decays do not contribute to a peak because the pion mass was assigned to the proton track.

Vertices corresponding to photon conversions, $\Kshort$-meson and $\Lambda$-baryon decays were vetoed by restrictions on the invariant mass of the tracks belonging to the secondary vertex. In the case of $\Lambda$ baryons, the mass was recalculated by assigning the proton mass to the track with the higher momentum, which gave the correct assignment in over $97\%$ of such decays. If the invariant mass lay within $\pm35 \MeV$ of the $\Kshort$ mass ($497.5\MeV$) or $\pm15 \MeV$ of the $\Lambda$ mass ($1116\MeV$), the vertex was vetoed. Track combinations with invariant masses below $310 \MeV$ were removed to eliminate photon conversions.

\FloatBarrier

\begin{table} [!htb]
\begin{center}
\caption{Yields of secondary vertices in data in various regions of the ID. The radius and $z$ of these regions are defined in Table~\ref{tab:TabPurity}.\label{tab:VtxYield}}
\vskip 0.1cm
\begin{tabular}{|l|r|}
\hline
\multicolumn{2}{|c|}{Pixel barrel region} \\
\hline \hline
Layer & Yield \\
\hline
Beam pipe & 1327622  \\
1st pixel layer & 1614080 \\
2nd pixel layer  & 548313  \\
3rd pixel layer  & 316376  \\
\hline \hline
\multicolumn{2}{|c|}{SCT barrel region} \\
\hline \hline
Layer & Yield \\
\hline
Pixel support frame & 170003 \\
Pixel support tube & 69682 \\
1st SCT layer & 134299  \\
2nd SCT layer & 28457 \\
\hline \hline
\multicolumn{2}{|c|}{Pixel forward region} \\
\hline \hline
Layer &  Yield \\
\hline
1st pixel disk  & 53448 \\
2nd pixel disk & 30119 \\
3rd pixel disk & 13694 \\
\hline
\end{tabular}
\end{center}
\end{table}

\begin{figure}[!thb]
\centering
\includegraphics[width=1.0\textwidth]{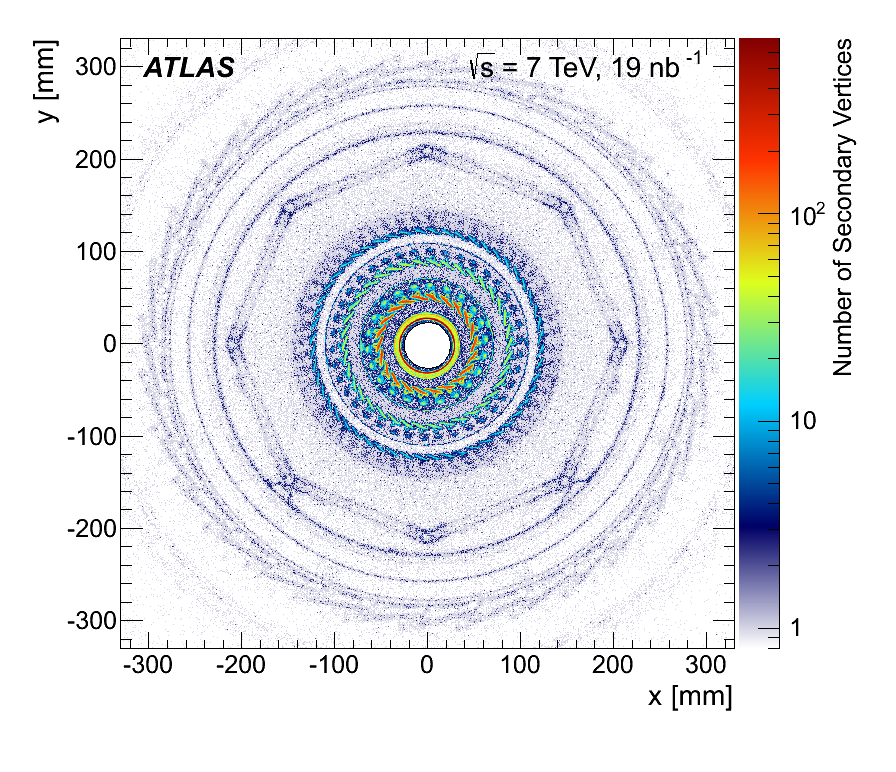} \\
\caption{\label{fig:SVXY} Number of selected secondary vertices in data in the \emph{x--y} plane of the ID. Only vertices with $|\emph{z}| < 400$~mm are considered. For presentation purposes, the background inside the beam pipe is masked out.}
\end{figure}

\begin{figure}[!thb]
\centering
\includegraphics[width=1.0\textwidth]{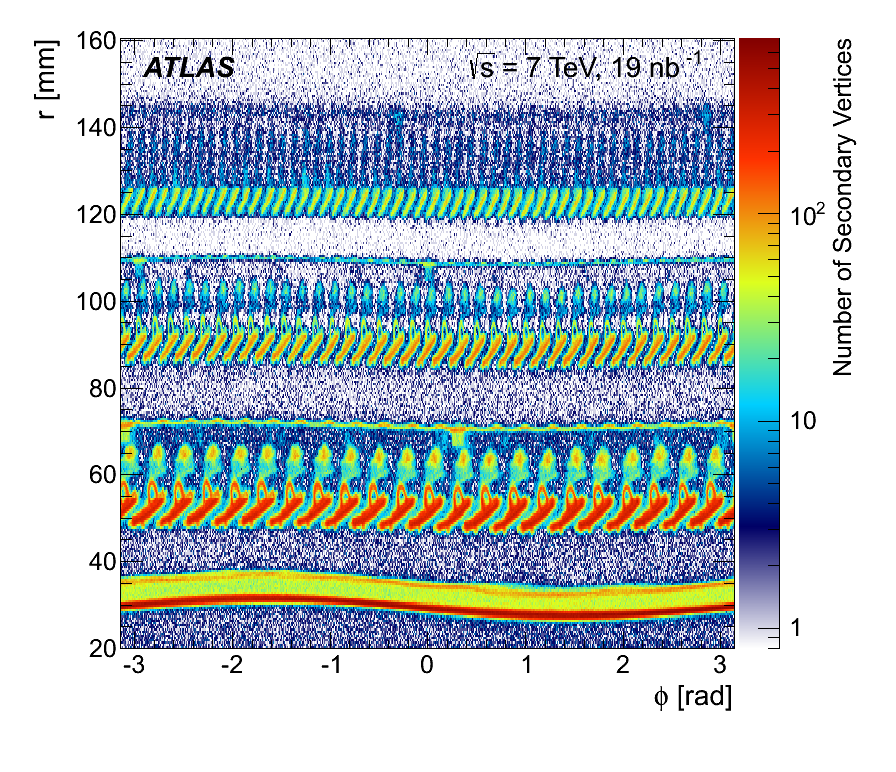} \\
\caption{\label{fig:SVRPhi} Number of selected secondary vertices in data in the $r$--$\phi$~plane of the ID. Only vertices with $|\emph{z}| < 400$~mm are considered.}
\end{figure}

\begin{figure}[!thb]
\centering
\includegraphics[width=1.0\textwidth]{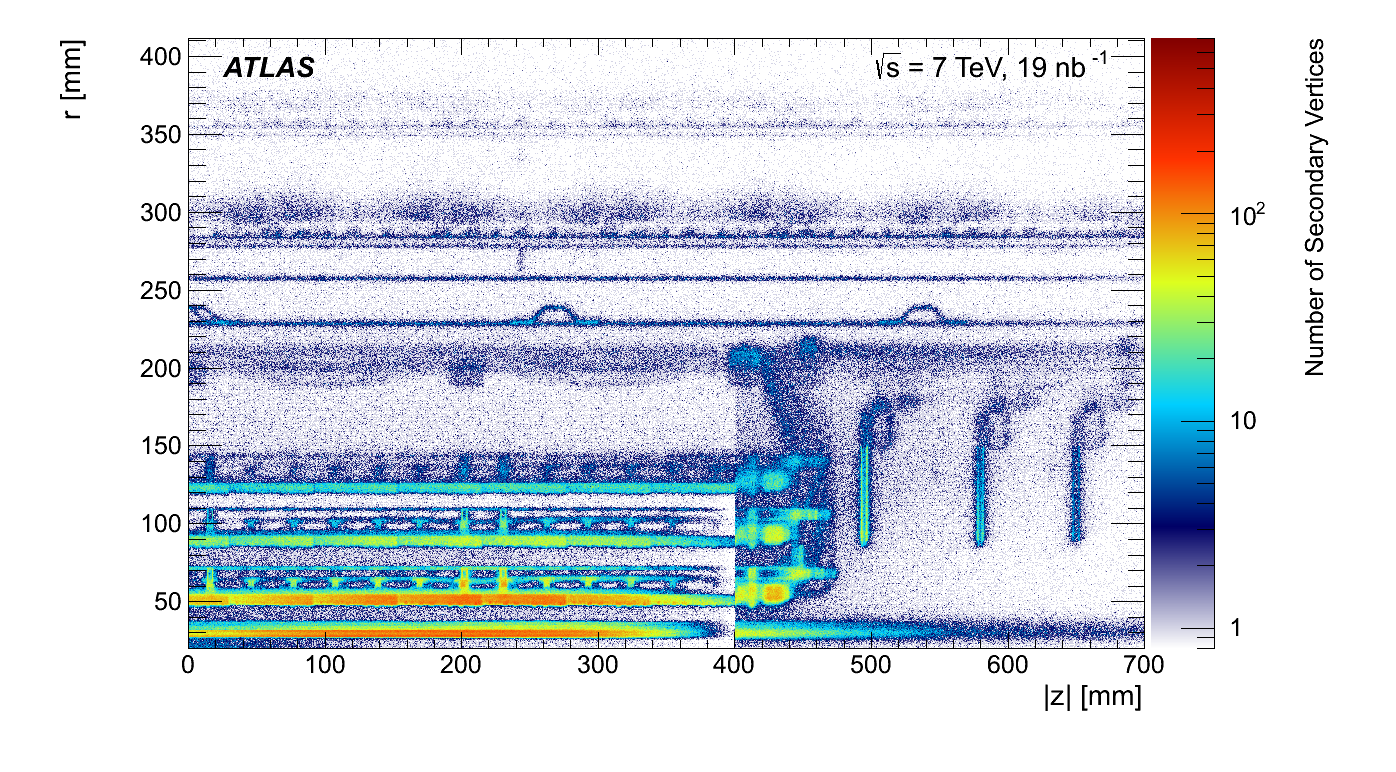} \\
\caption{\label{fig:SVRZEC} Number of selected secondary vertices in data in the $r$--$z$~plane of the ID.}
\end{figure}

\noindent
The vertex yields in data after these selection criteria are shown in Table~\ref{tab:VtxYield}. High-resolution images, in data, of the secondary-vertex positions in the detector are shown in Figures~\ref{fig:SVXY},~\ref{fig:SVRPhi} and \ref{fig:SVRZEC}. In Figure~\ref{fig:SVXY} the $y$-position is shown versus the $x$-position for vertices in the pixel barrel region, $|z|<400$~mm. The regions with higher density than their surroundings are, going from the innermost to the outermost: the beam pipe, the first, second and third pixel layers, the pixel support frame (the octagon), the pixel support tube, the SCT thermal shield, the first SCT layer and at the very edge with few vertices is the second SCT layer. Vertices further out than the second SCT layer are expected to have worse resolution and purity due to fewer silicon hits on the associated tracks; these layers are therefore disregarded.

In Figure~\ref{fig:SVRPhi}, clear details of the composition of the beam pipe are visible, e.g. the increase in vertex density at $\emph{r}\!\sim \!29$ mm indicates the layer made out of beryllium. Moreover, the fine structure of the modules in the three pixel barrel layers is visible. The wave-like behaviour in the beam pipe and the three pixel layers is due to the fact that the layers are not centred around [0,0]. The average displacements in \emph{x} and \emph{y} are listed in Table~\ref{tab:Center}.  The decrease in the number of vertices between the pixel barrel and the pixel end-caps, at $|z|\!\sim \!400$~mm in Figure~\ref{fig:SVRZEC}, is due to differences in the silicon-hit based fake-removal criteria in these two regions. The dense cluster of vertices at $|z|\!\sim$400--470~mm and $50 < r < 155$~mm represents services at the end of the pixel barrel. The pixel end-cap disks are at $z > 495$~mm, and $r$ between 80 and 175~mm.

\section{Qualitative comparison of data to simulation}
\label{sec:QualComp}

\begin{figure}[!htb]
\centering
\includegraphics[width=0.7\textwidth]{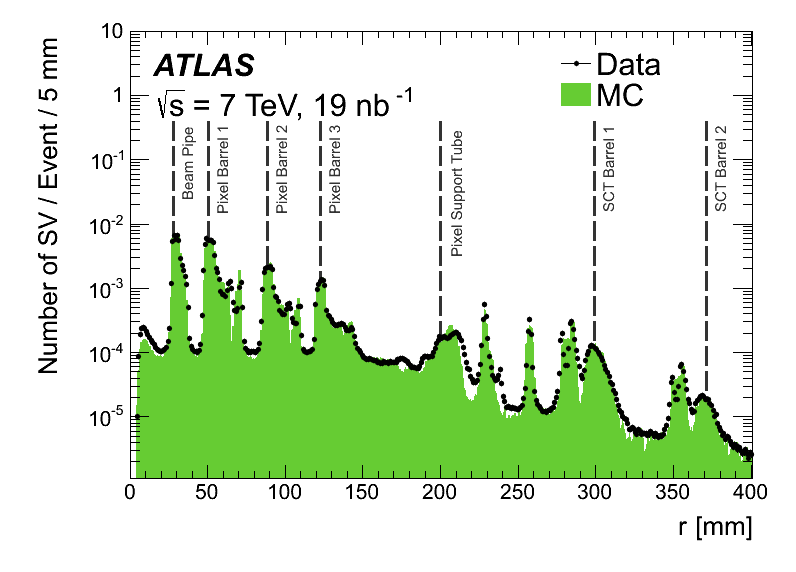} \\
\caption{\label{fig:RZSpectra} Number of selected secondary vertices per event in data (points) and  MC simulation (solid histogram), as a function of \emph{r} with $|z| < 700$ mm.}

\end{figure}

Vertex yields as a function of radius are shown in Figure~\ref{fig:RZSpectra}. Each material layer is clearly visible as indicated by the increase in vertex density at specific values of \emph{r}. Agreement between data and simulation is good, especially for the detector layers. However, there are small disagreements arising from simplifications made in the geometry model for some of the structures, e.g., the pixel support tube at $r\! \sim \!200$~mm, and others which appear as sharp peaks, e.g., \emph{r} around 70 and 105~mm. There are also some disagreements in the air gaps between detector layers.

\begin{figure}[htbp]
\begin{center}
\begin{tabular}{cc}
\includegraphics[width=0.5\textwidth]{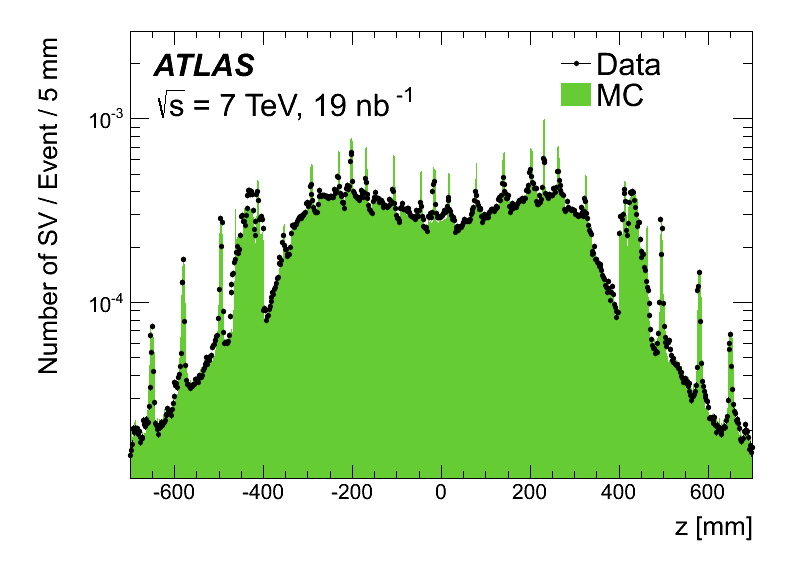} & \includegraphics[width=0.5\textwidth]{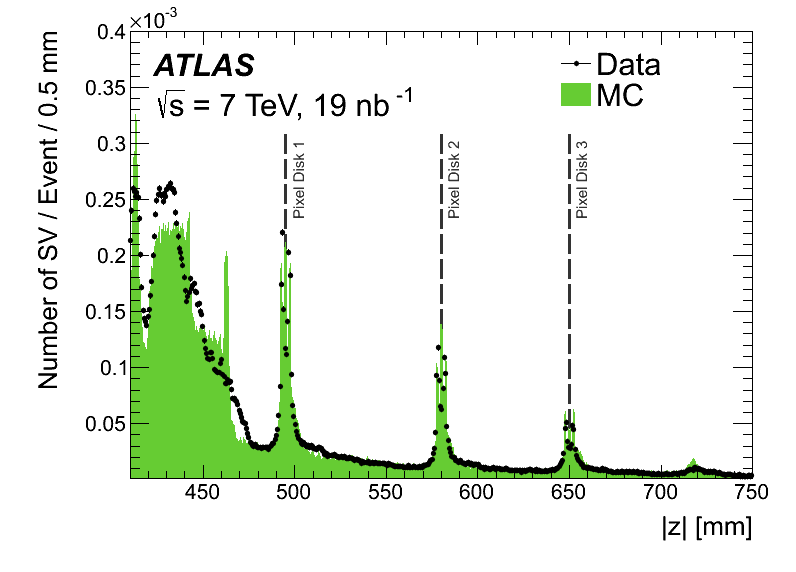} \\
(a) & (b) \\
\end{tabular}
\caption{Number of selected secondary vertices per event in data (points) and  MC simulation (solid histogram), as a function of ~\emph{z}, (a) for $|z|<700$~mm, and (b)  for $|z|>410$~mm, where vertices with $\emph{r} < 50$ mm and $\emph{r} > 155$ mm are excluded to highlight the pixel end-cap region.
\label{fig:ZPixEnd}}
\end{center}
\end{figure}

The distribution of secondary vertices for the full \emph{z}-range is shown in Figure~\ref{fig:ZPixEnd}(a), and the pixel end-cap region is shown separately in Figure~\ref{fig:ZPixEnd}(b). The peaks seen in simulation for $|z| < 300$ mm are due to simplifications of the support structures. The sudden drop at $|z| > 300 $ mm until around $400$~mm is due to differences in the silicon-hit based fake-removal criteria when going from the barrel to the end-cap region. The end of the pixel barrel and the three pixel end-cap disks correspond to the four spikes at $|z| > 400$~mm. To make the comparisons in the pixel end-cap region, vertices with $\emph{r} < 50$~mm and $\emph{r} > 155$~mm are excluded to eliminate regions such as the beam pipe and various support structures. The three pixel disks are clearly visible at $z\sim$ 495, 580 and 650 mm. Some discrepancies are observed in the dense cluster of vertices in Figure~\ref{fig:ZPixEnd}(b) with $|z|<470$~mm. This is a very complex region to simulate since it includes infrastructure for the cooling pipes and services for the pixel detector. Furthermore, simplifications in the geometry model lead to the sharp spikes in the disks themselves. Generally the simulation agrees well with data.

\FloatBarrier

\subsection{Displacement of material layers}
\label{sec:ShiftDetLay}

As reported previously~\cite{VJain}, the positions of the beam pipe and the three pixel layers are not centred around [$x=0$,~$y=0$] in the detector, as concluded from the sinusoidal behaviour in Figure~\ref{fig:SVRPhi}. Sine functions were fit to the $\phi$ dependence in each radial interval to estimate the displacement from the origin in \emph{x} and \emph{y}. Their amplitudes are reported in Table~\ref{tab:Center}; uncertainties coming from the fitting procedure are $\lesssim 0.02$~mm. The shift in simulation agrees well with data for the beam pipe, and the small discrepancy for the pixel layers has no impact on the quantitative results presented later in this paper.

\begin{figure}[!thb]
\begin{center}
\includegraphics[width=0.9\textwidth]{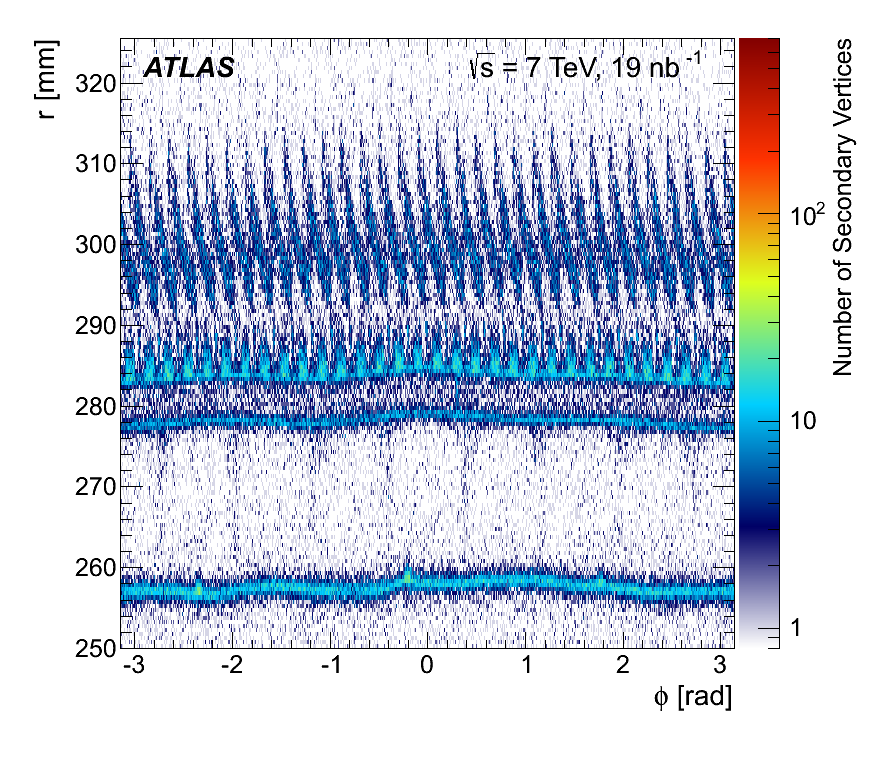}
\caption{\label{fig:SCTDef} Number of selected secondary vertices in data in the \emph{r}--$\phi$ plane of the ID. The radial range corresponds to the SCT inner thermal shield and the first SCT barrel layer.  }
\end{center}
\end{figure}

Figure~\ref{fig:SCTDef} shows the $r$--$\phi$ distribution of selected secondary vertices for the radial region corresponding to the first SCT layer ($r$:~276--320~mm).  The SCT inner thermal shield is at $r\approx 257$~mm. Several deformations and shifts are visible in the thermal shield, e.g. most clearly the "bumps" at various $\phi$ positions ($-1.8$, $-0.2$ and +1.0) with a size of ~0.5~mm; shifts are of order 0.5~mm to 1~mm. In simulated events, these structures are flat along the $\phi$-axis.

\begin{table}[!htb]
\begin{center}
\caption{\label{tab:Center}The displacement, \emph{x} and \emph{y}, in mm from the origin for the first four material layers of the ID. Displacements seen in simulated events are compared with data. The radius and $z$ of the various layers are defined in Table~\ref{tab:TabPurity}.}
\begin{tabular}{| l |c | c | c | c | c|}
\hline
 &  \multicolumn{2}{c|}{Data} & \multicolumn{2}{c|}{MC simulation}  \\
Layer &  Disp. \emph{x} & Disp. \emph{y} & Disp. \emph{x} & Disp. \emph{y} \\
\hline \hline
Beam pipe &  --0.21 & --1.90 &  --0.21 & --1.93 \\
1st pixel layer &  --0.32 &  --0.50 & --0.23 & --0.32 \\
2nd pixel layer &  --0.34 & --0.46 & --0.25 & --0.35 \\
3rd pixel layer &  --0.15 & --0.16 & --0.31 & --0.18 \\
\hline
\end{tabular}
\end{center}
\end{table}

\FloatBarrier

\subsection{Pixel and SCT detector modules in their local coordinate frames}

\label{sec:Modules}
The detector modules in the barrel overlap in \emph{r}--$\phi$ to give complete coverage, as can be seen in Figure~\ref{fig:SVXY}. To study details of individual modules, the positions of the reconstructed vertices were transformed from the global ATLAS reference frame to the module-specific local coordinate frame.\footnote{Local-\emph{X} is along the global $\phi$ direction, local-\emph{Y} is along the stave (global $z$), and local-\emph{Z} is along the global radial direction} These transformations were performed for vertices in the first five silicon layers of the ID barrel. Due to the decreasing number of vertices with increasing \emph{r} the visible details of the pixel modules are degraded when going from the first to the third layer, and similarly for the SCT layers. Hence for qualitative comparisons, only modules in the first pixel and SCT layers are considered. These are shown in Figures~\ref{fig:ModulesPixel1} and \ref{fig:ModulesSCT1} respectively.

Due to the excellent radial vertex resolution it is possible to specifically select vertices originating in the silicon sensor element in the pixel and SCT modules; the thickness of this element in these two kinds of modules is $250~\mu$m and $ 285~\mu$m respectively. These vertices are then used in Section~\ref{sec:comp_modelling} to explore kinematic details of the modelling of hadronic interactions in the silicon.

\subsubsection{Pixel modules}
\label{sec:ModulesPixel}

\begin{figure}[!tbh]
\begin{center}
\begin{tabular}{cc}
\includegraphics[width=0.49\textwidth]{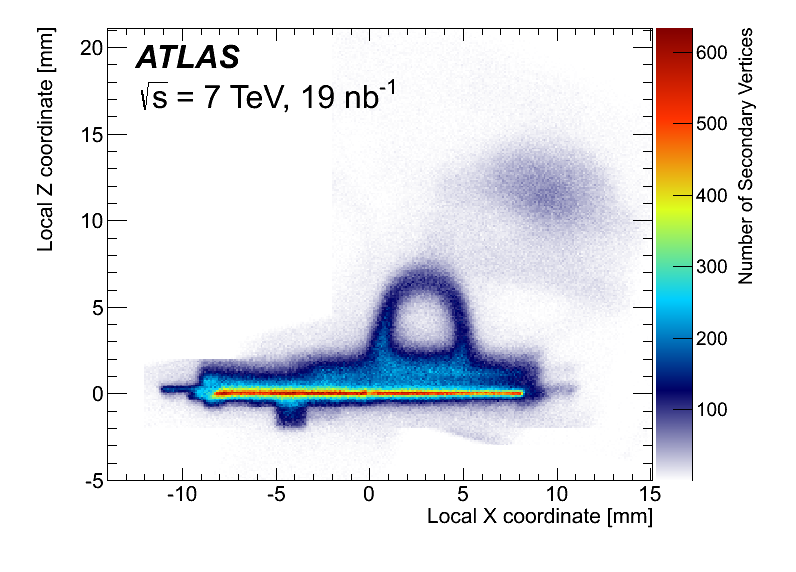} & \includegraphics[width=0.49\textwidth]{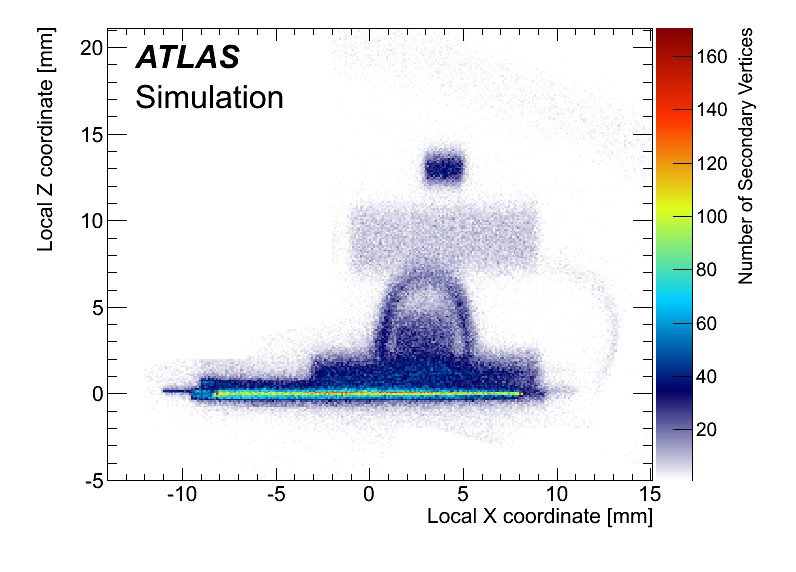} \\
(a) & (b) \\
\end{tabular}
\caption{\label{fig:ModulesPixel1} Number of selected secondary vertices in the local coordinate frame of the pixel module, for (a) data, and (b) MC simulation, in the first barrel layer. The data sample has more events than the simulated sample.}
\end{center}
\end{figure}

Components of the pixel module such as the active silicon sensor element (local-\emph{X} within $\pm 9$~mm and local-$Z\!\sim \!0$~mm), the cooling-fluid pipe (seen as the half-circle on top of the module), and cables and supports (the rectangular area above the cooling pipe in simulation, and above and to the right in data) are clearly visible in Figure~\ref{fig:ModulesPixel1}. 

\FloatBarrier

\subsubsection{SCT modules}
\label{sec:ModulesSCT}

\begin{figure}[!tbh]
\begin{center}
\begin{tabular}{cc}
\includegraphics[width=0.49\textwidth]{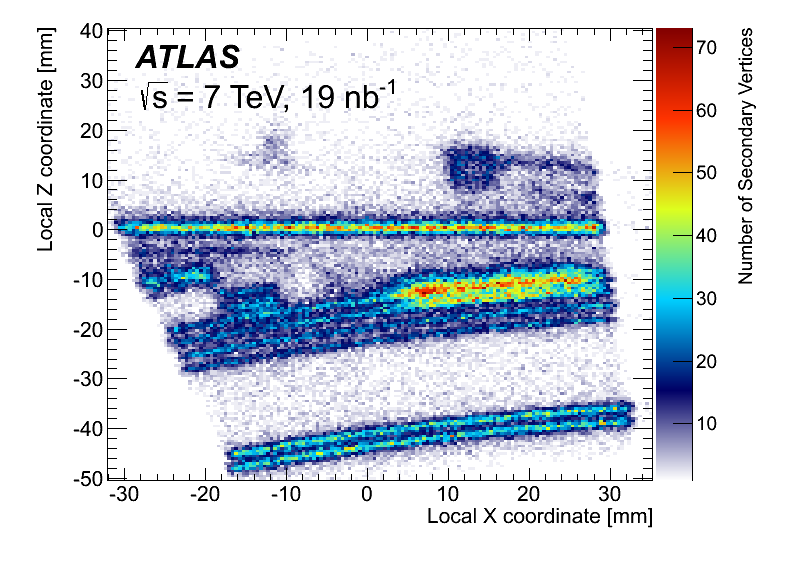} & \includegraphics[width=0.49\textwidth]{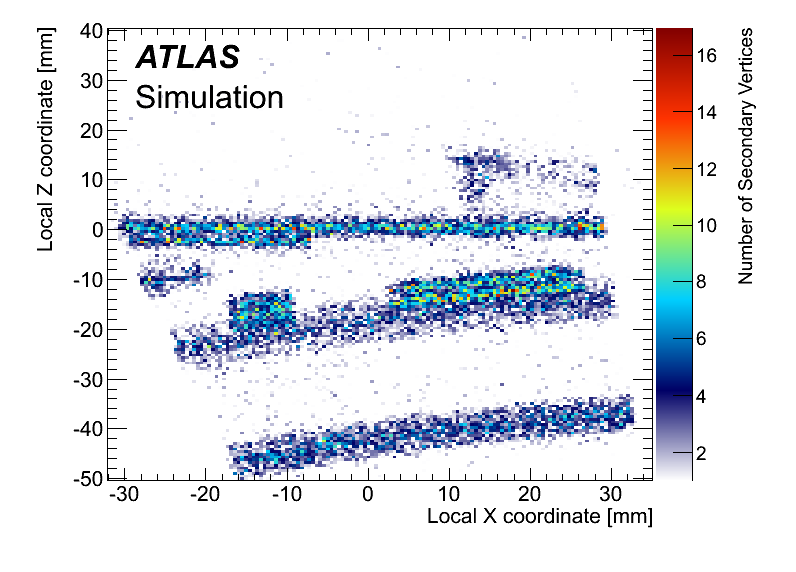} \\
(a) & (b) \\
\end{tabular}
\caption{\label{fig:ModulesSCT1} Number of selected secondary vertices in the local coordinate frame of the SCT module, for (a) data and (b) MC simulation, in the first barrel layer. The data sample has more events than the simulated sample.}
\end{center}
\end{figure}

Figure~\ref{fig:ModulesSCT1} shows a detailed view of a module in the first SCT barrel layer in data and in simulation. The horizontal section around local-\emph{Z} of 0~mm spanning local-\emph{X} within $\pm 30$~mm is the sensor element, and the structure right above it (with local-$X$ in the range [10,30]~mm) is the cooling-fluid pipe. The band with local-\emph{Z} in the range [$-30,-20$]~mm is a support structure, and the band at the bottom of the plot is the SCT thermal shield.

\subsection{Kinematic characteristics of secondary vertices}
\label{sec:comp_modelling}

In addition to comparing the location of secondary vertices in data and simulation, their kinematic characteristics are compared. This allows a study of models of hadronic interactions of low- and medium-energy hadrons used in GEANT4. Kinematic variables that are mainly dependent on details of the hadronic interactions are compared, e.g., the total momentum of tracks emerging from the secondary vertex, the fraction of the total momentum carried by the highest-momentum secondary track, and the opening angle between the tracks.

\begin{figure}[!tbh]
\begin{center}
\begin{tabular}{cc}
\includegraphics[width=0.49\textwidth]{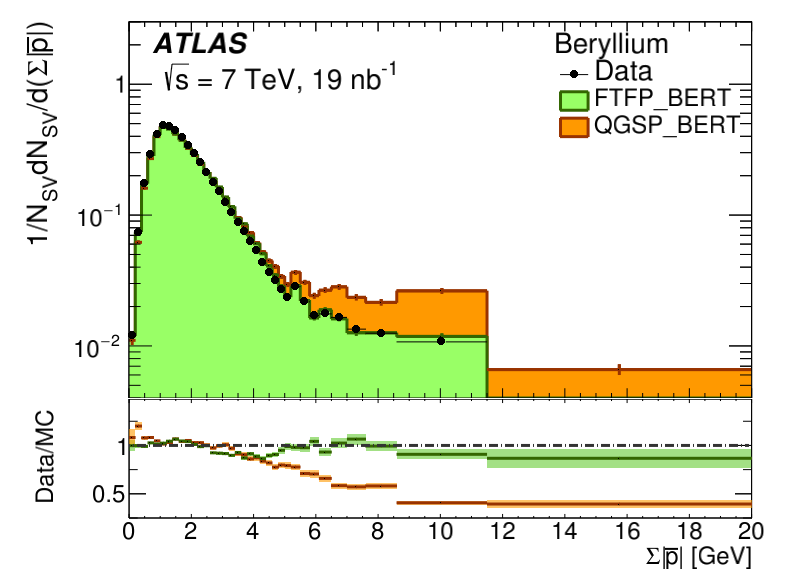} & \includegraphics[width=0.49\textwidth]{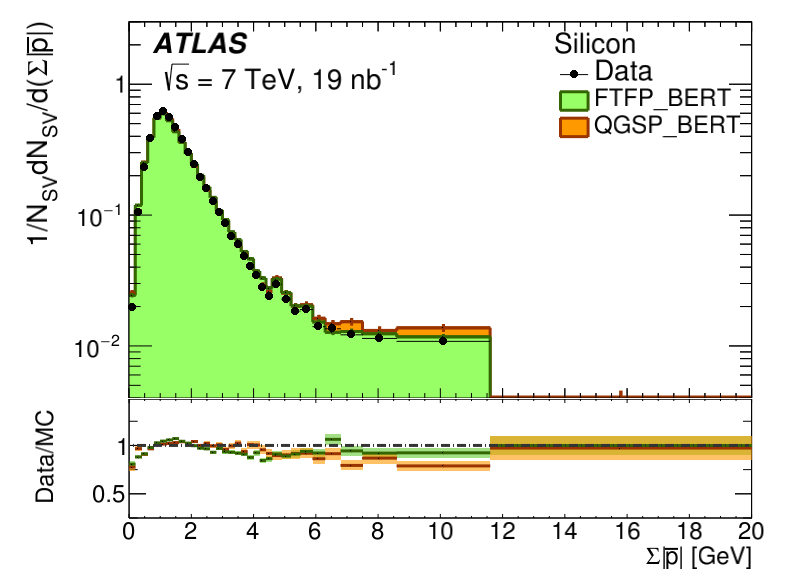} \\
(a) & (b) \\
\end{tabular}
\caption{\label{fig:G4SecMom} The total (scalar) sum of the momentum, $\Sigma |\bar{p}|$, of secondary particles from hadronic interactions found in (a) the beryllium part of the beam pipe, and (b) the silicon element in the pixel modules. The data (points) are compared to simulation using two models of hadronic interactions (solid histograms). The number of selected secondary vertices is $N_{\mathrm{SV}}$.}
\end{center}
\end{figure}

These variables are explored by isolating structures within the ID that are composed of single elements, the $800~\mu$m beryllium layer in the beam pipe and the $250~\mu$m silicon sensor within the pixel modules in the barrel. Two different models of hadronic interactions (FTFP\_BERT and QGSP\_BERT)~\cite{Models} are compared with data. Figure~\ref{fig:G4SecMom} shows the total (scalar) momentum sum of the tracks emerging from a secondary vertex in data compared with the two interaction models in (a) beryllium, and (b) silicon;  the FTFP\_BERT model agrees better with data, and is used in the nominal MC simulation sample.

\FloatBarrier

\section{Systematic uncertainties}
\label{sec:syst}
\subsection{Tracking efficiency}

The main source of systematic uncertainty in the reconstruction efficiency of charged hadrons is the uncertainty in the description of material in the ID~\cite{PrimTrk}. The overall scale of the reconstruction efficiency of primary tracks in data is well modelled in simulation, implying that the total amount of material in the ID is well understood. However, discrepancies in the location of the material can impact the reconstruction efficiency of tracks arising from vertices far from the PV.

Decays of $\Kshort$-mesons are used to make a comparison of track reconstruction efficiency in data and simulation as a function of vertex position. These decays are a source of charged pions that are independent of the hadronic interaction rate, yet probe the material in a manner similar to tracks from a secondary interaction vertex. Pions produced in these decays also have large impact parameters relative to the PV. The momentum spectrum of $\Kshort$ candidates in simulation agrees with data, hence a comparison of their yields (in data and simulation) as a function of decay distance probes the efficiency of reconstructing secondary tracks. $\Kshort$ yields are determined by fitting the invariant mass of two tracks of opposite charge with a signal and a background function, where the signal is a sum of two Gaussian functions (with a common mean) and the background is represented by a first-order polynomial.

The hadronic-interaction analysis provides differential measurements of the material budget, layer by layer. Comparing the reconstructed $\Kshort$ yield in data to that in simulation in different radial intervals corresponding to the material layers eliminates any direct dependency of the track reconstruction efficiency on the material in that layer, since this efficiency is dependent only on the amount of material \emph{outside} the given material layer. Intervals were selected so that the tracks from the $\Kshort$ decays in the selected radial range passed through the same amount of material as tracks from hadronic interactions: 0--48~mm, 48--86~mm, 86--120~mm, 120--276~mm, 276--347~mm, and 347--500~mm.

\begin{figure}[!htb]
\centering
\includegraphics[width=0.7\textwidth]{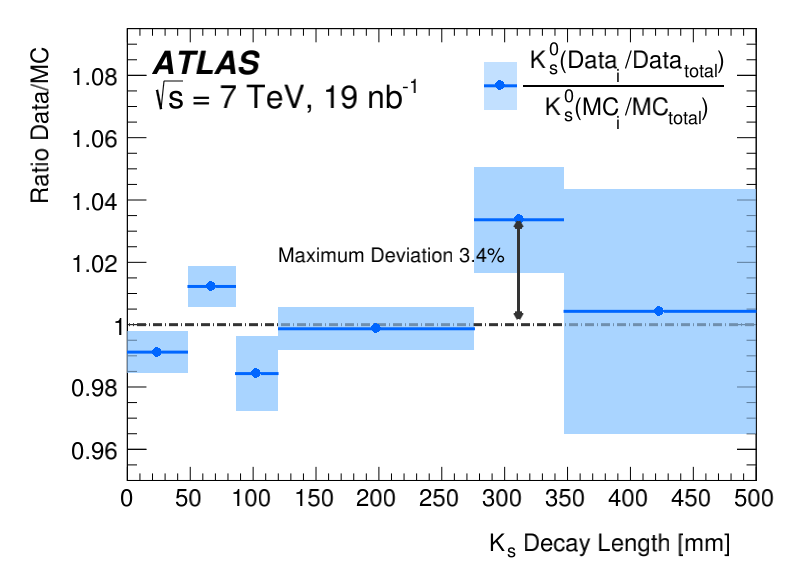}
\caption{A double ratio of $\Kshort$ yields at different decay distances. The bands are statistical uncertainties from the fits to obtain K$_s$ yields. Details are given in the text. }
\label{fig:KsRatio}
\end{figure}

To account for different numbers of $\Kshort$-mesons produced in data and simulation, yields in various radial bins in data and simulation were normalized separately to the total yields in the respective samples. A double ratio of these normalized yields in data and simulation was constructed, and is shown in Figure~\ref{fig:KsRatio}. A deviation from unity would imply a difference between track reconstruction efficiencies in data and simulation. The largest deviation is about 3.4\%. 

The uncertainty arising from the production of $\Kshort$-mesons in secondary interactions or decays of heavier hadrons was estimated to be 0.3\%. This was added linearly to the uncertainty determined above to give a total systematic uncertainty of 3.7\% for a two-track vertex. This implies an uncertainty of 1.85\% per track. Since the maximum deviation from unity was taken as a measure of the systematic uncertainty, it is likely to be conservative.

Since about 10\% of secondary vertices from hadronic interactions have more than two associated tracks, an overall systematic uncertainty was derived by randomly removing $1.85\%$ of the tracks during vertex reconstruction. In vertices with more than two tracks, removing one of the tracks still allows the vertex to be reconstructed. The reduction in the number of secondary vertices was $3.4\%$, which was taken as the systematic uncertainty on the ratio of hadronic interaction vertex yields in data and simulation.

\FloatBarrier

\subsection{Modelling of hadronic interactions in simulation}
\label{sec:modelling}

A quantitative comparison of vertex yields in data and simulation is sensitive to the modelling of low- and medium-energy hadronic interactions in GEANT4. If incorrect hadronic cross-sections were to be used in simulation, that would lead to an incorrect prediction for the number of secondary vertices. To study this, hadronic interaction cross-sections used in the nominal simulation sample (FTFP\_BERT model) were compared with two of its variants, FTFP\_BERT\_chipsXS and FTFP\_BERT\_gheishaXS~\cite{Models}. These cross-sections were generated separately for charged pion, kaon, and proton interactions with beryllium, carbon and silicon for a variety of incident energies. Using the differences between these cross-sections, and the numbers of interactions in various elements due to these hadrons, the following systematic uncertainties were assigned: $^{+0.2\%}_{-0.4\%}$ for the beam pipe (which is mainly beryllium), $\pm 0.1\%$ for the pixel support frame and tube (mainly made of carbon), and $^{+0.9\%}_{-3.2\%}$ for all other material layers (assumed to be silicon and carbon).

In addition, mismodelling of secondary hadronic interactions in GEANT4 can lead to differences in the distributions of kinematic variables between data and simulation. Several variables were studied: invariant mass and $\Sigma(\pt)$ of the tracks emerging from the secondary vertex, $(\pt(\mathrm{max}) - \pt(\mathrm{min}))/\Sigma(\pt)$, where min (max) are the minimum (maximum) \pt~of secondary tracks, and the polar angle of the momentum vector of the secondary vertex, which was determined using the momenta of the secondary tracks. Ratios of yields in data to those in simulation were determined for different intervals in these variables, and based on these studies, layer-by-layer systematic uncertainties were estimated. These were combined in quadrature with previously determined uncertainties in the interaction cross-sections used in GEANT4. The total uncertainties are shown in Table~\ref{tab:G4Systematics}; they are applied to the ratios of secondary-vertex yields in data to those in simulation.

\begingroup
\renewcommand{\arraystretch}{1.3}
\begin{table}[!htb]
\begin{center}
\caption{\label{tab:G4Systematics} Systematic uncertainties due to mismodelling of hadronic interactions in GEANT4 in various regions of the ID. These uncertainties are applied to the ratios of secondary-vertex yields in data to those in simulation. The radius and $z$ of these regions are defined in Table~\ref{tab:TabPurity}.}
\vskip 0.1cm
\begin{tabular}{| l |c | }
\hline
\multicolumn{2}{|c|}{Pixel barrel region} \\
\hline
\hline
Layer & Syst. Uncert. \\
\hline 
Beam pipe & $^{+0.2\%}_{-0.4\%}$ \\[0.15cm]
1st pixel layer & $^{+1.6\%}_{-3.5\%}$\\[0.15cm]
2nd pixel layer & $^{+2.7\%}_{-4.1\%}$\\[0.15cm]
3rd pixel layer & $^{+2.8\%}_{-4.1\%}$\\[0.15cm]
\hline \hline
\multicolumn{2}{|c|}{SCT barrel region} \\
\hline \hline
Layer & Syst. Uncert. \\
\hline 
Pixel support frame & $^{+2.9\%}_{-2.9\%}$ \\[0.15cm]
Pixel support tube & $^{+3.4\%}_{-3.4\%}$\\[0.15cm]
1st SCT layer & $^{+3.0\%}_{-4.3\%}$ \\[0.15cm]
2nd SCT layer & $^{+3.6\%}_{-4.7\%}$\\[0.15cm]
\hline \hline
\multicolumn{2}{|c|}{Pixel forward region} \\
\hline \hline
Layer & Syst. Uncert. \\
\hline 
1st pixel disk & $^{+1.7\%}_{-3.5\%}$\\[0.15cm]
2nd pixel disk & $^{+1.4\%}_{-3.4\%}$\\[0.15cm]
3rd pixel disk & $^{+1.2\%}_{-3.3\%}$\\[0.15cm] 
\hline

\end{tabular}
\end{center}
\end{table}
\endgroup

\FloatBarrier

\subsection{Vertex reconstruction}

Systematic uncertainties arising during the vertex selection step could come from non-optimal modelling of the $\chi^{2}$ requirement on the vertex fit, and the merging of nearby vertices. Data and simulated samples were re-analysed using different selection criteria for these variables, and the relative differences in the behaviour of these samples were used to assign a systematic uncertainty of 1\%.

Studies on simulated events showed that once restrictions were placed on the invariant mass of tracks emerging from secondary vertices to remove contamination from photon conversions, $\Kshort$ and $\Lambda$ decays, the remaining vertices were mainly due to true hadronic interactions or were fake, i.e., composed of random tracks that do not originate from a common point, and consequently have large values of $\chi^2$/dof. Any residual contamination from conversions and $\Kshort$ and $\Lambda$ decays was found to be less than 1\%, and was ignored.

However, an additional systematic uncertainty can arise from mismodelling the rate of fake secondary vertices. At the beam pipe, the purity is 82.5\%, i.e., 16.5\% of the reconstructed vertices are expected to be fake. In some of the other material layers, the fraction of fake vertices is higher (Table~\ref{tab:TabPurity}). Vertices with $10 < \chi^2/{\mathrm {dof}} < 50$ were chosen to study this effect. The ratio of reconstructed vertices in data and simulation, with these poor values of $\chi^2$/dof, was studied in three intervals of vertex radius, $0-100$~mm, $100-180$~mm and $\ge 180$~mm, and was found to differ from unity by 8\%, 8\% and 13\% respectively. Using these numbers in conjunction with the expected fake rate at various material layers led to systematic uncertainties shown in Table~\ref{tab:SystPurity}. These uncertainties are applied to the ratios of secondary-vertex yields in data to those in simulation.

\begin{table}
\begin{center}
\caption{\label{tab:SystPurity} Systematic uncertainties due to fake vertices in various regions of the ID. These are applied to the ratios of secondary-vertex yields in data to those in simulation. The radius and $z$ of these regions are defined in Table~\ref{tab:TabPurity}.}
\vskip 0.1cm
\begin{tabular}{|l|c|}
\hline
\multicolumn{2}{|c|}{Pixel barrel region} \\
\hline \hline
Layer & Syst. Uncert. \\
\hline
Beam pipe & $1.4\%$ \\
1st pixel layer & $2.0\%$ \\
2nd pixel layer & $1.6\%$ \\
3rd pixel layer & $2.6\%$ \\
\hline \hline
\multicolumn{2}{|c|}{SCT barrel region} \\
\hline \hline
Layer & Syst. Uncert. \\
\hline
Pixel support frame & $3.6\%$ \\
Pixel support tube  & $1.7 \%$ \\
1st SCT layer & $2.8\%$ \\
2nd SCT layer & $5.0\%$ \\
\hline \hline
\multicolumn{2}{|c|}{Pixel forward region} \\
\hline \hline
Layer & Syst. Uncert. \\
\hline
1st pixel disk  & $3.4\%$ \\
2nd pixel disk & $3.0\%$ \\
3rd pixel disk & $3.5\%$ \\
\hline
\end{tabular}
\end{center}
\end{table}

\subsection{Other sources}
\label{sec:otherSources}

In this analysis, only the number of primary tracks in simulation was corrected to match data, and because hadronic interactions caused by secondary particles and neutral hadrons were not explicitly accounted for, a systematic uncertainty due to this source was investigated. Correction factors for primary and secondary tracks were studied separately, where the latter were selected by requiring $|d_0|$ to be $>5$~mm, and an average was determined based on estimates from simulation for the fraction of interactions that were due to secondary particles. An uncertainty of 1\% was assigned due to this source. Since the production rate of neutral hadrons is related to the rate for charged hadrons via isospin symmetry, and corrections were made for charged particles, any residual uncertainties were assumed to be negligible.

As discussed in Section~\ref{sec:Samples}, a requirement on the track multiplicity at the PV was made in order to enhance non-diffractive events in data. If the distributions were to be different in data and simulation, this could introduce an uncertainty. This was tested by increasing the requirement on the number of tracks in the PV and noting the reduction in the number of vertices, separately in data and simulation samples. A difference of 0.3\% was seen between the two samples, and this was taken as a measure of the systematic uncertainty from this source.

Another uncertainty could arise if the primary-particle composition was incorrectly generated in PYTHIA8~\cite{PYTHIA}. This event generator was tuned using data collected by experiments at LEP and SLAC, which had dedicated particle identification sub-detectors. Also, a comparison of PYTHIA8 and  EPOS~\cite{epos} samples (for 13 TeV \emph{pp} collisions) showed good agreement in the generated fractions of charged hadrons such as pions, kaons and protons. Hence, the systematic uncertainty from this source was assumed to be small and was neglected.

\subsection{Total systematic uncertainty}
The systematic uncertainties on the ratios of yields in data to those in simulation are listed in Table~\ref{tab:Systematics}. They are assumed to be uncorrelated and combined in quadrature.

\begin{table}[!htb]
\begin{center}
\caption{\label{tab:Systematics} Summary table of all systematic uncertainties in the vertex yield comparisons.}
\vskip 0.1cm
\begin{tabular}{| c |c | }
\hline
Source & Syst. Uncert. \\
\hline \hline
Track reconstruction efficiency & 3.4\% \\[0.1cm]\hline
Modelling of hadronic interactions & {$-4.7$\%} -- {+3.6\%} \\ \hline
Vertex reconstruction & 1.0\% \\ \hline
Fake vertices & 1.4--5.0\% \\ \hline
Track multiplicity correction for MC simulation & 1.0\% \\ \hline
Primary track multiplicity cut & 0.3\% \\

\hline
\end{tabular}
\end{center}
\end{table}

\FloatBarrier

\section{Results}
\label{sec:Results}

The ratios of yields of reconstructed vertices per event in data to those in simulation are used to make a quantitative estimate of how well the detector material is modelled in the simulation. In this approach, various acceptances and reconstruction efficiencies largely cancel, and any residual differences between data and simulation are assigned as systematic uncertainties on the ratio of yields; these uncertainties are listed in Table~\ref{tab:Systematics}.

\begin{figure}[!htb]
\centering
\includegraphics[width=0.7\textwidth]{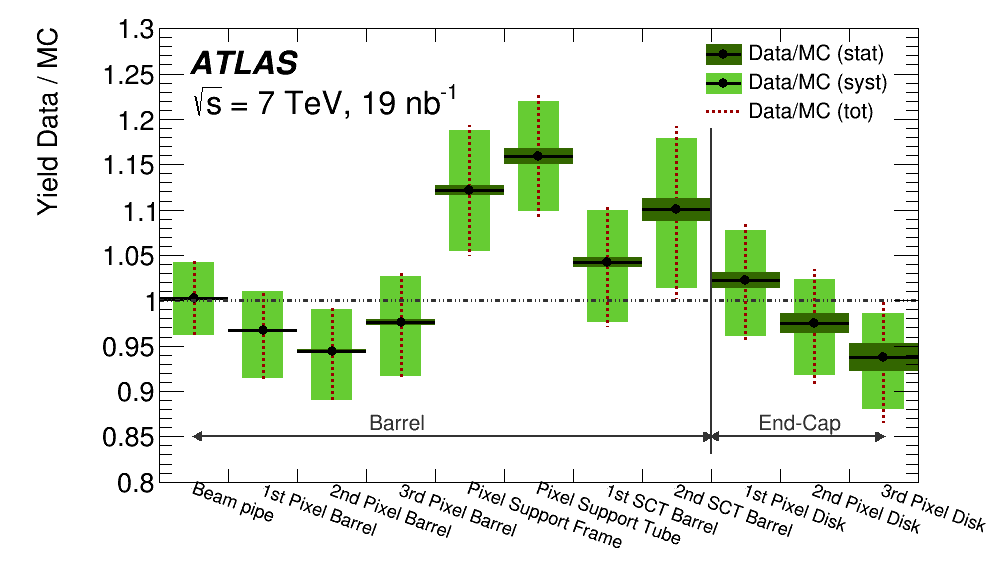}
\caption{Ratios of yields of secondary vertices in data to those in MC simulation for the barrel and end-cap regions. The total uncertainties (dotted lines) include both statistical (dark shading) and systematic uncertainties (light shading) in the vertex yields.}
\label{fig:Yield}
\end{figure}

The ratios of yields for the beam pipe, the three pixel layers, the pixel support frame and tube, the first two SCT layers and the three pixel end-cap disks are shown in Table~\ref{tab:SystTotal} and Figure~\ref{fig:Yield}. All secondary-vertex quality cuts were applied before vertex yields were calculated. The yields in the simulated sample were multiplied by the correction factors to account for different numbers of primary tracks in data and simulation. Simulated events were also weighted to match the \emph{z}-coordinate of the primary-vertex spectrum in data.  The total uncertainty is dominated by systematic effects, discussed in Section~\ref{sec:syst}. 

The beam-pipe region shows good agreement between data and simulation; in addition, the yields in the 800~$\mu$m beryllium layer in the beam pipe agree well, where the ratio of data to simulation is $1.010\pm0.002\stat \pm 0.040\syst$.\footnote{Material in the as-built beam pipe~\cite{detector} is known to 3\%} In the three pixel barrel layers the simulation overestimates the material by about $1\sigma$ while in the two SCT barrel layers the material in the data is larger than in the simulation by also about $1\sigma$. The first two pixel disks agree well within uncertainties, while the third disk is low by about $1\sigma$. An  excess of material in the radial regions 180--220~mm (pixel support frame) and 226--240~mm (pixel support tube) is observed in data as compared to the ATLAS detector model; these two structures combined constitute less than 1\% of the total mass of the ID. These discrepancies are understood to be due to missing components in the description of these support structures, which will be included in newer versions of the ATLAS detector geometry used in simulation.

\begingroup
\renewcommand{\arraystretch}{1.3}
\begin{table}
\begin{center}
\caption{\label{tab:SystTotal} Ratios of yields of secondary vertices in data to those in MC simulation for various regions in the ID. The first uncertainty is statistical and the second systematic. The radius and $z$ of these regions are defined in Table~\ref{tab:TabPurity}.}
\vskip 0.1cm
\begin{tabular}{|l|l|}
\hline
\multicolumn{2}{|c|}{Pixel barrel region} \\
\hline \hline
Layer & Yield Data/MC \\
\hline
Beam pipe & $1.002\pm0.002\pm0.040$ \\[0.15cm]
1st pixel layer & $0.967\pm0.001~_{-0.053}^{+0.043}$ \\[0.2cm]
2nd pixel layer & $0.944\pm0.002~_{-0.054}^{+0.046}$ \\[0.2cm]
3rd pixel layer  & $0.976\pm0.003~_{-0.060}^{+0.051}$ \\[0.2cm]
\hline \hline
\multicolumn{2}{|c|}{SCT barrel region } \\
\hline \hline
Layer & Yield Data/MC \\
\hline
Pixel support frame  & $1.121\pm0.006\pm0.067$ \\[0.2cm]
Pixel support tube  & $1.159\pm0.009\pm0.061$ \\[0.2cm]
1st SCT layer & $1.042\pm0.006~_{-0.066}^{+0.058}$ \\[0.2cm]
2nd SCT layer  & $1.100\pm0.013~_{-0.086}^{+0.079}$ \\[0.2cm]
\hline \hline
\multicolumn{2}{|c|}{Pixel forward region} \\
\hline \hline
Layer & Yield Data/MC \\
\hline
1st pixel disk & $1.023\pm0.009~_{-0.062}^{+0.054}$ \\[0.2cm]
2nd pixel disk & $0.975\pm0.011~_{-0.057}^{+0.049}$ \\[0.2cm]
3rd pixel disk & $0.937\pm0.015~_{-0.057}^{+0.049}$ \\[0.2cm]
\hline
\end{tabular}
\end{center}
\end{table}
\endgroup 

\FloatBarrier

Table~\ref{tab:ModYield} lists the ratios of yields in the silicon element of the detector modules in the first five silicon layers in the barrel using the local coordinate transformations discussed in Section~\ref{sec:Modules}. The agreement is within uncertainties. Since the thickness of the silicon element is well-known, these results demonstrate that experimental uncertainties are well understood.

\begingroup
\renewcommand{\arraystretch}{1.3}
\begin{table}[h]
\begin{center}
\caption{\label{tab:ModYield} The ratios of vertex yields in data to those in MC simulation for the silicon sensor element in the pixel and SCT modules, measured using the local coordinate transformations. The first uncertainty is statistical and the second systematic. The radius and $z$ of the various regions are defined in Table~\ref{tab:TabPurity}.}
\begin{tabular}{|l|c|}
\hline
\multicolumn{2}{|c|}{Pixel barrel region} \\
\hline
\hline
Module location & Yield Data/MC\\
\hline
1st pixel layer & $0.999\pm0.003~_{-0.054}^{+0.045}$ \\[0.2cm]
2nd pixel layer & $0.945\pm0.004~_{-0.054}^{+0.045}$ \\[0.2cm]
3rd pixel layer & $0.957\pm0.006~_{-0.058}^{+0.050}$ \\[0.2cm]
\hline \hline
\multicolumn{2}{|c|}{SCT barrel region} \\
\hline \hline
Module location & Yield Data/MC\\
\hline
1st SCT layer & $1.029\pm0.011_{-0.065}^{+0.057}$ \\[0.2cm]
2nd SCT layer & $1.035\pm0.030_{-0.081}^{+0.074}$ \\[0.2cm]
\hline
\end{tabular}
\end{center}
\end{table}
\endgroup

\section{Conclusions}
\label{sec:Conclusions}

This paper presents an updated study of the material in the ATLAS ID using secondary hadronic interactions in $19$~nb$^{-1}$ of $\sqrt{s}$~=~7~\TeV~LHC \emph{pp} collision events collected with a minimum-bias trigger. The technique described here exploits primary particle interactions with ID material to
reconstruct secondary vertices from the outgoing tracks. Since the incident primary particles have low to medium energies, the outgoing tracks tend to have large opening angles between them, thereby improving spatial resolutions. A major improvement is due to second-pass track reconstruction that increases the efficiency of reconstructing secondary tracks, especially those arising from vertices far from the primary vertex.

These secondary vertices are reconstructed using an inclusive vertex-finding and fitting algorithm, and they have good spatial resolution in both the longitudinal and transverse directions. The resulting resolutions are significantly better than the resolution of vertices produced by photon conversions, which are routinely used for material estimation. This leads to a precise `radiography' of the as-built ID and facilitates comparison with the implementation of the detector geometry in simulation.

Many sources of systematic uncertainty have been investigated, viz., secondary track and vertex reconstruction efficiencies, $\pt$ and $\eta$ distributions of primary particles that interact in the ID, and the accuracy of hadronic interaction modelling in GEANT4. Experimental systematic uncertainties, i.e., those arising from track and vertex reconstruction, are estimated from data in this analysis. Differences between data and simulation in the $\pt$ and $\eta$ distributions of the primary tracks are accounted for via a reweighting procedure. Uncertainties in the modelling of hadronic interactions in GEANT4 are studied by comparing various models.

Due to the increase in vertex reconstruction efficiency from second-pass track reconstruction, this study explores a large fiducial volume (0.7~m$^3$). The second barrel SCT layer and the pixel end-cap disks have been investigated for the first time with this technique. Moreover, this paper includes `radiography' images of the SCT modules, which had not been studied previously. Agreement between data and simulation is good to within the experimental uncertainties, which are about 5\%.

\section*{Acknowledgements}


We thank CERN for the very successful operation of the LHC, as well as the
support staff from our institutions without whom ATLAS could not be
operated efficiently.

We acknowledge the support of ANPCyT, Argentina; YerPhI, Armenia; ARC, Australia; BMWFW and FWF, Austria; ANAS, Azerbaijan; SSTC, Belarus; CNPq and FAPESP, Brazil; NSERC, NRC and CFI, Canada; CERN; CONICYT, Chile; CAS, MOST and NSFC, China; COLCIENCIAS, Colombia; MSMT CR, MPO CR and VSC CR, Czech Republic; DNRF and DNSRC, Denmark; IN2P3-CNRS, CEA-DSM/IRFU, France; GNSF, Georgia; BMBF, HGF, and MPG, Germany; GSRT, Greece; RGC, Hong Kong SAR, China; ISF, I-CORE and Benoziyo Center, Israel; INFN, Italy; MEXT and JSPS, Japan; CNRST, Morocco; FOM and NWO, Netherlands; RCN, Norway; MNiSW and NCN, Poland; FCT, Portugal; MNE/IFA, Romania; MES of Russia and NRC KI, Russian Federation; JINR; MESTD, Serbia; MSSR, Slovakia; ARRS and MIZ\v{S}, Slovenia; DST/NRF, South Africa; MINECO, Spain; SRC and Wallenberg Foundation, Sweden; SERI, SNSF and Cantons of Bern and Geneva, Switzerland; MOST, Taiwan; TAEK, Turkey; STFC, United Kingdom; DOE and NSF, United States of America. In addition, individual groups and members have received support from BCKDF, the Canada Council, CANARIE, CRC, Compute Canada, FQRNT, and the Ontario Innovation Trust, Canada; EPLANET, ERC, FP7, Horizon 2020 and Marie Sk{\l}odowska-Curie Actions, European Union; Investissements d'Avenir Labex and Idex, ANR, R{\'e}gion Auvergne and Fondation Partager le Savoir, France; DFG and AvH Foundation, Germany; Herakleitos, Thales and Aristeia programmes co-financed by EU-ESF and the Greek NSRF; BSF, GIF and Minerva, Israel; BRF, Norway; Generalitat de Catalunya, Generalitat Valenciana, Spain; the Royal Society and Leverhulme Trust, United Kingdom.

The crucial computing support from all WLCG partners is acknowledged gratefully, in particular from CERN, the ATLAS Tier-1 facilities at TRIUMF (Canada), NDGF (Denmark, Norway, Sweden), CC-IN2P3 (France), KIT/GridKA (Germany), INFN-CNAF (Italy), NL-T1 (Netherlands), PIC (Spain), ASGC (Taiwan), RAL (UK) and BNL (USA), the Tier-2 facilities worldwide and large non-WLCG resource providers. Major contributors of computing resources are listed in Ref.~\cite{ATL-GEN-PUB-2016-002}.


\printbibliography

\clearpage
\begin{flushleft}
{\Large The ATLAS Collaboration}

\bigskip

M.~Aaboud$^{\rm 135d}$,
G.~Aad$^{\rm 86}$,
B.~Abbott$^{\rm 113}$,
J.~Abdallah$^{\rm 64}$,
O.~Abdinov$^{\rm 12}$,
B.~Abeloos$^{\rm 117}$,
R.~Aben$^{\rm 107}$,
O.S.~AbouZeid$^{\rm 137}$,
N.L.~Abraham$^{\rm 149}$,
H.~Abramowicz$^{\rm 153}$
% ...
\bigskip
\\
% ...
$^{12}$ Institute of Physics, Azerbaijan Academy of Sciences, Baku, Azerbaijan\\
% ...
$^{64}$ University of Iowa, Iowa City IA, United States of America\\
% ...
$^{86}$ CPPM, Aix-Marseille Universit{\'e} and CNRS/IN2P3, Marseille, France\\
% ...
$^{107}$ Nikhef National Institute for Subatomic Physics and University of Amsterdam, Amsterdam, Netherlands\\
% ...
$^{113}$ Homer L. Dodge Department of Physics and Astronomy, University of Oklahoma, Norman OK, United States of America\\
% ...
$^{117}$ LAL, Univ. Paris-Sud, CNRS/IN2P3, Universit{\'e} Paris-Saclay, Orsay, France\\
% ...
$^{135}$ $^{(a)}$ Facult{\'e} des Sciences Ain Chock, R{\'e}seau Universitaire de Physique des Hautes Energies - Universit{\'e} Hassan II, Casablanca; $^{(b)}$ Centre National de l'Energie des Sciences Techniques Nucleaires, Rabat; $^{(c)}$ Facult{\'e} des Sciences Semlalia, Universit{\'e} Cadi Ayyad, LPHEA-Marrakech; $^{(d)}$ Facult{\'e} des Sciences, Universit{\'e} Mohamed Premier and LPTPM, Oujda; $^{(e)}$ Facult{\'e} des sciences, Universit{\'e} Mohammed V, Rabat, Morocco\\
% ...
$^{137}$ Santa Cruz Institute for Particle Physics, University of California Santa Cruz, Santa Cruz CA, United States of America\\
% ...
$^{149}$ Department of Physics and Astronomy, University of Sussex, Brighton, United Kingdom\\
% ...
$^{153}$ Raymond and Beverly Sackler School of Physics and Astronomy, Tel Aviv University, Tel Aviv, Israel\\
% ...
$^{a}$ Also at Department of Physics, King's College London, London, United Kingdom\\
$^{b}$ Also at Institute of Physics, Azerbaijan Academy of Sciences, Baku, Azerbaijan\\
$^{c}$ Also at Novosibirsk State University, Novosibirsk, Russia\\
$^{d}$ Also at TRIUMF, Vancouver BC, Canada\\
$^{e}$ Also at Department of Physics {\&} Astronomy, University of Louisville, Louisville, KY, United States of America\\
% ...
\end{flushleft}



\end{document}